\documentclass[prd,twocolumn,amsfonts,amssymb]{revtex4}

\usepackage{graphicx}

\usepackage{bm}

\newcommand{\beq}{\begin{equation}}
\newcommand{\eeq}{\end{equation}}
\newcommand{\beqs}{\begin{eqnarray}}
\newcommand{\eeqs}{\end{eqnarray}}
\newcommand{\lsim}{\mathrel{\raisebox{-
.6ex}{$\stackrel{\textstyle<}{\sim}$}}}

\begin{document}

\title{Improved Constraints on Sterile Neutrinos in the MeV to GeV Mass Range} 

\author{D. A. Bryman$^{a,b}$ and R. Shrock$^c$}

\affiliation{(a) \ Department of Physics and Astronomy, University of British 
Columbia, Vancouver, British Columbia, V6T 1Z1, Canada}

\affiliation{(b) \ TRIUMF, 4004 Wesbrook Mall, Vancouver, British Columbia, V6T
  2A3, Canada} 

\affiliation{(c) \ C. N. Yang Institute for Theoretical Physics and 
Department of Physics and Astronomy, \\
Stony Brook University, Stony Brook, NY 11794, USA }

\begin{abstract}

  Improved upper bounds are presented on the coupling $|U_{e4}|^2$ of an
  electron to a sterile neutrino $\nu_4$ from analyses of data on nuclear and
  particle decays, including superallowed nuclear beta decays, the ratios
  $R^{(\pi)}_{e/\mu}=BR(\pi^+ \to e^+ \nu_e)/BR(\pi^+ \to \mu^+ \nu_\mu)$,
  $R^{(K)}_{e/\mu}$, $R^{(D_s)}_{e/\tau}$, and $B^+_{e 2}$ decay, covering the
  mass range from MeV to GeV. 

\end{abstract}

\maketitle

Neutrino oscillations and hence neutrino masses and lepton mixing have
been established and are of great importance as physics beyond the
original Standard Model (SM). Most oscillation experiments with solar,
atmospheric, accelerator, and reactor (anti)neutrinos can be explained 
within the minimal framework of three neutrino mass eigenstates with
values of $\Delta m^2_{ij} =
m_{\nu_i}^2 - m_{\nu_j}^2$ given approximately by $\Delta m^2_{21} =
0.74 \times 10^{-4}$ eV$^2$ and $|\Delta m^2_{32}| = 2.5 \times 10^{-3}$
eV$^2$, with normal mass ordering $m_{\nu_3} > m_{\nu_2}$ favored; furthermore,
the lepton mixing angles $\theta_{23}$, $\theta_{12}$, and $\theta_{13}$ 
have been measured, with a tentative indication of a nonzero value of 
the CP-violating quantity $\sin(\delta_{CP})$ 
\cite{pdg2018}-\cite{cggnov}.

In addition to the three known neutrino mass eigenstates, there could be
others, which would necessarily be primarily electroweak-singlets (sterile)
\cite{st}. Indeed, sterile neutrinos are present in many ultraviolet (UV)
extensions of the SM. Whether sterile neutrinos exist in nature is one of the
most outstanding questions in particle physics, and therefore, improved
constraints on their couplings are of fundamental and far-reaching importance.
Taking account of the possibility of sterile neutrinos, the neutrino
interaction eigenstates $\nu_\ell$ would be given by
\beq
\nu_\ell = \sum_{i=1}^{3+n_s} U_{\ell i} \, \nu_i \ , 
\label{nuell}
\eeq
where $\ell=e, \ \mu, \ \tau$; $n_s$ denotes the number of sterile
neutrinos; and $U$ is the lepton mixing matrix \cite{anom}. 

Here we obtain improved upper limits on $|U_{e i}|^2$ for a sterile neutrino
$\nu_i$ in a wide range of masses from the MeV to GeV scale and point out new
experiments that would be worthwhile and could yield further improvements.  
For simplicity, we assume one heavy neutrino, $n_s=1$, with $i=4$; it is
straightforward to generalize to $n_s \ge 2$. Since a $\nu_4$ in this mass
range decays, it is not excluded by the cosmological upper limit on the sum of
effectively stable neutrinos, $\sum_i m_{\nu_i} \lsim 0.12$ eV
\cite{planck2018}. Such a $\nu_4$ is subject to a number of constraints from
cosmology (e.g., \cite{cosm}); however, since these depend on assumptions about
the early universe, we choose here to focus on direct laboratory
bounds. Constraints from the non-observation of neutrinoless double beta decay
are satisfied by assuming that $\nu_4$ is a Dirac neutrino \cite{dirac}.  
Since sterile neutrinos violate the conditions for the
diagonality of the weak neutral current \cite{leeshrock77,sv80}, $\nu_4$ has
invisible tree-level decays of the form $\nu_4 \to \nu_j \bar \nu_i \nu_i$
where $1 \le i, j \le 3$ with model-dependent invisible branching
ratios. Because our bounds are purely kinematic, they are complementary to
bounds from searches for neutrino decays, which involve model-dependent
assumptions on branching ratios into visible versus invisible final states.

We first obtain improved upper bounds on $|U_{e 4}|^2$ from nuclear beta
decay data. The emission of a $\nu_4$ via lepton mixing in nuclear beta
decay has several effects, including producing a kink in the Kurie plot and
reducing the decay rate \cite{shrock80}. For the nuclear beta decays $(Z,A) \to
(Z + 1,A) + e^- + \bar\nu_e$ or $(Z,A) \to (Z-1,A)+e^+ + \nu_e$ into a set of
neutrino mass eigenstates $\nu_i \in \nu_e$, $i=1,2,3$ of negligibly small
masses, plus a $\nu_4$ of of non-negligible mass, the differential decay rate
is
\begin{widetext}
\beq
\frac{dN}{dE} = C \bigg [ (1-|U_{e 4}|^2)p E (E_0-E)^2 
 + |U_{e 4}|^2 p E (E_0-E)\Big [(E_0-E)^2-m_{\nu_4}^2 \Big ]^{1/2} \, 
\theta(E_0-E-m_{\nu_4}) \bigg ] \ ,
\label{nucldecrate}
\eeq
\end{widetext}
where $p\equiv |{\mathbf p}|$ and $E$ denote the 3-momentum and (total) energy
of the outgoing $e^\pm$, $E_0$ denotes its maximum energy for the SM case, 
the Heaviside $\theta$ function is defined as
$\theta(x)=1$ for $x > 0$ and $\theta(x)=0$ for $x \le 0$, and $C = G_F^2
|V_{ud}|^2 F_F |{\cal M}|^2/(2\pi^3)$, where ${\cal M}$ denotes the nuclear
transition matrix element, $V$ is the Cabibbo-Kobayashi-Maskawa (CKM) quark
mixing matrix, and $F_F$ is the Fermi function.  
Early bounds on $|U_{e4}|^2$ were set from
searches for kinks in Kurie plots in \cite{shrock80} and analyses of 
particle decays \cite{shrock81a}-\cite{shrock_vpi}, and from 
dedicated experiments. For example, a search for kinks in the Kurie
plot in ${}^{20}$F beta decay reported in Ref. \cite{deutsch1990}
yielded an upper bound on $|U_{e4}|^2$ decreasing from $5.9 \times
10^{-3}$ for $m_{\nu_4}=0.4$ MeV to $1.8 \times 10^{-3}$ for
$m_{\nu_4}=2.8$ MeV.  (Some recent reviews of searches for sterile neutrinos
include \cite{kusenkorev}-\cite{batell2018}.)

In addition to kink searches, a powerful method to set constraints on massive
neutrino emission, via lepton mixing, in nuclear beta decays is to analyze the
decay rates. Since, in general, the heavy neutrino would also be emitted in
$\mu$ decay, the measurement of the $\mu$ lifetime performed assuming the SM
would yield an apparent ($app$) value of the Fermi constant, denoted
$G_{F,app}$, that would be smaller than the true value, $G_F$, given at tree
level by $G_F/\sqrt{2} = g^2/(8m_W^2)$, where $g$ is the SU(2) gauge coupling
\cite{shrock81a,shrock81b,shrock_vpi}.  To avoid this complication, 
the ratios of rates of different nuclear beta decays are compared.

The integration of $dN/dE$ over $E$ gives the kinematic rate factor $f$. The
combination of this with the half-life for the nuclear beta decay, $t \equiv
t_{1/2}$, yields the product $ft$. Incorporation of nuclear and radiative
corrections yields the corrected $ft$ value for a given decay, denoted ${\cal
  F}t$.  Conventionally, analyses of the ${\cal F}t$ values for the most
precisely measured superallowed $0^+ \to 0^+$ nuclear beta decays have been
used, in conjunction with the value of $G_{F,app}$ from $\mu$ decay,
to infer a value of the weak mixing matrix element, $|V_{ud}|$
\cite{ht75}-\cite{vudapp}.  A first step in these analyses has been to
establish the mutual consistency of the ${\cal F}t$ values for these
superallowed $0^+ \to 0^+$ decays. Since the emission of a $\nu_4$ with 
mass of a few MeV would have a different effect on the kinematic 
functions and integrated rates for nuclear beta decays with different 
$Q$ (energy release) values, it would upset this mutual consistency. 

\begin{figure}[hbt]
\centering
\includegraphics[height=12cm,angle=0,scale=0.5]{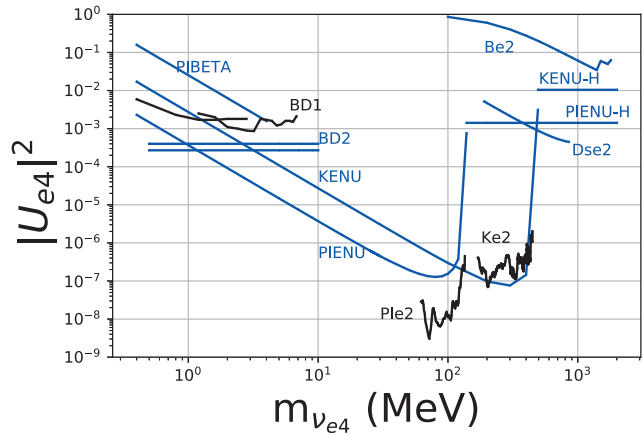}
\caption{90 \% C.L. upper limits on $|U_{e 4}|^2$ {\it vs.}
$m_{\nu_4}$ from various sources:  PIBETA, pion beta decay\cite{DB_RS};
BD1, previous limits from beta decay \cite{deutsch1990};
BD2, beta decay with the two dashed horizonal lines based on our analysis using
\cite{hardy_towner2018} and \cite{ramsey_musolf}; PIENU and PIENU-H, the ratio
$\frac{BR(\pi^+ \to e^+\nu_e)}{BR(\pi^+ \to \mu^+ \nu_\mu)}$ in
the kinematically allowed and forbidden regions for $\nu_4$ emission;
$Pie2$, $\pi^+ \to e^+ \nu_{e4}$ peak search \cite{pienu2018};
KENU and KENU-H, the ratio
$\frac{BR(K^+ \to e^+\nu_e)}{BR(K^+ \to \mu^+ \nu_\mu)}$ in
the kinematically allowed and forbidden regions for $\nu_4$ emission;
$Ke2$, $K^+ \to  e^+ \nu_{e4}$ peak search \cite{na62_2018};
$Dse2$, $D_s^+ \to e^+ \nu_{e4}$; and, $Be2$, $B^+ \to e^+ \nu_{e4}$.}
\label{Ue4_figure}
\end{figure}

Hence, from this mutual agreement of ${\cal F}t$ values, an upper limit on
$|U_{e4}|^2$ can be derived for values of $m_{\nu_4}$ such that a $\nu_4$ could
be emitted in some of these superallowed decays.  Ref. \cite{hardy_towner1990}
obtained upper bounds on $|U_{e4}|^2$ ranging from $10^{-2}$ down to $2 \times
10^{-3}$ for $m_{\nu_4}$ from 0.5 to 2 MeV, while Ref.  \cite{deutsch1990}
obtained the limits $|U_{e4}|^2 < 1 \times10^{-3}$ to $|U_{e4}|^2 < 2 \times
10^{-3}$ for $m_{\nu_4}$ from 1 to 7 MeV.  The maximum $Q$ value in the current
set of 14 superallowed $0^+ \to 0^+$ beta decays used for the ${\cal F}t$ fit
in \cite{hardy_towner2015,hardy_towner2018} is 9.4 MeV (for ${}^{74}$Rb).  A
measure of the mutual agreement is the precision with which $|V_{ud}|^2$ is
determined, so a reduction in the fractional uncertainty of the value of
$|V_{ud}|^2$ results in an improved upper limit on
$|U_{e4}|^2$. Ref. \cite{hardy_towner1990} obtained $|V_{ud}|=0.9740 \pm
0.001$. The recent analyses in \cite{hardy_towner2018} and \cite{ramsey_musolf}
obtain $|V_{ud}| = 0.97420(21)$ and $|V_{ud}| = 0.97370(14)$, respectively
\cite{ckmu}. Applying these factors of improvement from \cite{hardy_towner2018}
and \cite{ramsey_musolf} to the previous bounds in \cite{hardy_towner1990},
improved upper bounds are obtained as 
\beq
|U_{e4}|^2 < 4 \times 10^{-4}  
\label{ue4sq}
\eeq
and
\beq
|U_{e4}|^2 <  2.7 \times 10^{-4} 
\label{ue4sq_rm}
\eeq
for $\nu_4$ masses in the range $0.5 \ {\rm MeV} \lsim m_{\nu_4} < 9.4$ MeV,
indicated in Fig. \ref{Ue4_figure} as BD2. (These and other limits presented
are at the 90 \% confidence level.)

We next discuss upper bounds from two-body leptonic decays of charged
pseudoscalar mesons (generically denoted as $M^+$) \cite{shrock80,shrock81a}.
This method is quite powerful, because the signal is a monochromatic peak in
$dN/dp_\ell$ and for $M^+_{e2}$ decays, the strong helicity suppression in the
SM case is removed when a heavy neutrino is emitted.  The presence of a
massive $\nu_4$ also changes the ratio $BR(M^+\to e^+\nu_e)/BR(M^+ \to
\mu^+\nu_\mu)$ from its SM value,, and this was used to set further bounds
\cite{shrock80,shrock81a,bryman83b}. A number of dedicated experiments have 
been performed to search for a peak due to heavy neutrino emission and also to
measure $BR(M^+ \to e^+ \nu_e)/BR(M^+ \to \mu^+ \nu_\mu)$ with 
$\pi^+_{\ell 2}$, $K^+_{\ell  2}$, and $B^+_{\ell 2}$, where $\ell=e, \ \mu$ 
\cite{abela81}-\cite{triumf_pimu2}. 

In the SM with only the three known neutrinos with negligibly small masses, 
the ratio 
\beq
R^{(M)}_{\ell/\ell'} \equiv \frac{BR(M^+ \to \ell^+ \nu_\ell)}
                                 {BR(M^+ \to \ell'^+ \nu_{\ell'})}
\label{rmellellp}
\eeq
is given by 
\beqs
R^{(M)}_{\ell/\ell',SM} &=& \frac{m_\ell^2}{m_{\ell'}^2} \, \bigg [
\frac{1-\delta^{(M)}_\ell}
     {1-\delta^{(M)}_{\ell'}} \bigg ]^2 (1 + \delta_{RC}) \ ,
\label{rmsm}
\eeqs
where $\delta^{(M)}_\ell = m_\ell^2/m_M^2$ and 
$\delta_{RC}$ is the radiative correction (RC) \cite{earlyrad}-\cite{sdb}.

We denote the ratio of the experimental measurement
of $R^{(M)}_{\ell/\ell'}$ to the SM prediction as 
\beq
\bar R^{(M)}_{\ell/\ell'} \equiv \frac{R^{(M)}_{\ell/\ell'}}
                                        {R^{(M)}_{\ell/\ell',SM}} \ . 
\label{rmcal}
\eeq

The most precise measurement of $R^{(\pi)}_{e/\mu}$
is from the PIENU experiment at TRIUMF, with the result $R^{(\pi)}_{e/\mu} =
(1.2344 \pm 0.0023_{stat} \pm 0.0019_{syst} ) \times 10^{-4}$
\cite{pienu2015}. The resultant PDG world average is 
$R^{(\pi)}_{e/\mu} = (1.2327 \pm 0.0023) \times 10^{-4}$ \ \cite{pdg2018},  
in agreement with the SM prediction with RC,
$R^{(\pi)}_{e/\mu}=(1.2352 \pm 0.0002) \times 10^{-4}$ \  
\cite{ms93,cirigliano07,annrev2011}. 
Using the PDG value of $R^{(\pi)}_{e/\mu}$, one finds
\beq
\bar R^{(\pi)}_{e/\mu} = 0.9980 \pm 0.0019 \ . 
\label{rcal_pi_emu}
\eeq
The ratio $R^{(K)}_{e/\mu}$ has recently been measured by 
the NA62 experiment at CERN \cite{na62_kemu}, dominating the world average
\cite{pdg2018} 
\beq
R^{(K)}_{e/\mu} = (2.488 \pm 0.009) \times 10^{-5} \ . 
\label{rk_emu}
\eeq
The SM prediction with RC \cite{cirigliano07,sdb} is 
\beq
R^{(K)}_{e/\mu,SM} = (2.477 \pm 0.001) \times 10^{-5}  \ ,
\label{rk_emu_sm}
\eeq
resulting in 
\beq
\bar R^{(K)}_{e/\mu} = 1.0044 \pm 0.0037 \ . 
\label{rcal_k_emu}
\eeq

With emission of a heavy neutrino $\nu_4$, the ratio 
$R^{(M)}_{\ell/\ell',SM}$ for general $\ell, \ \ell'$ changes to 
\beqs
R^{(M)}_{\ell/\ell'} &=&
\Bigg [ \frac{[(1-|U_{\ell 4}|^2)\rho(\delta^{(M)}_\ell,0) + 
|U_{\ell 4}|^2 \rho(\delta^{(M)}_\ell,\delta^{(M)}_{\nu_4})}
{(1-|U_{\ell' 4}|^2)\rho(\delta^{(M)}_{\ell'},0) + 
|U_{\ell' 4}|^2 \rho(\delta^{(M)}_{\ell'},\delta^{(M)}_{\nu_4})} \Bigg ]
\times \cr\cr
&\times& (1+\delta_{RC}) \ , 
\label{rmgen}
\eeqs
where $\delta^{(M)}_{\nu_4} = m_{\nu_4}^2/m_M^2$, and  
the kinematic function $\rho(x,y)$ is \cite{shrock80,shrock81a}
\beqs
\rho(x,y) = [x+y-(x-y)^2][\lambda(1,x,y)]^{1/2}  \ , 
\label{rhom}
\eeqs
with
\beq
\lambda(z,x,y) = x^2+y^2+z^2-2(xy+yz+zx) \ .
\label{lam}
\eeq
Thus, in the SM case, $\rho(x,0)=x(1-x)^2$. Here and below, it is implicitly
understood that $\rho(\delta^{(M)}_\ell,\delta^{(M)}_{\nu_4}) = 0$ if
$m_{\nu_4} \ge m_M-m_\ell$, where the decay $M^+ \to \ell^+ \nu_4$
is kinematically forbidden.  We define 
\beq
\bar\rho(x,y) = \frac{\rho(x,y)}{\rho(x,0)} = \frac{\rho(x,y)}{x(1-x)^2} \ 
\label{rhobar}
\eeq
so 
\beqs
\bar R^{(M)}_{\ell/\ell'} = \frac{ 1-|U_{\ell 4}|^2 + |U_{\ell 4}|^2
\bar\rho(\delta^{(M)}_\ell,\delta^{(M)}_{\nu_4})}
{1-|U_{\ell' 4}|^2 + |U_{\ell' 4}|^2
\bar\rho(\delta^{(M)}_{\ell'},\delta^{(M)}_{\nu_4}) } \ .
\label{rr}
\eeqs

With no loss of generality, we order $\ell$ and $\ell'$ such that $m_{\ell'} >
m_\ell$ and define the mass intervals (i) $I^{(M)}_1: \ m_{\nu_4} <
m_M-m_{\ell'}$; (ii) $I^{(M)}_2: \ m_M-m_{\ell'} < m_{\nu_4} < m_M - m_\ell$;
and (iii) $I^{(M)}_3: \ m_{\nu_4} > m_M-m_\ell$. Thus, a $\nu_4$ with
$m_{\nu_4} \in I^{(M)}_1$ contributes to both $M^+_{\ell 2}$ and $M^+_{\ell'
  2}$ decays, while if $m_{\nu_4} \in I^{(M)}_2$, then $\nu_4$ contributes to
$M^+_{\ell 2}$, but not to $M^+_{\ell' 2}$ decay, and if $m_{\nu_4} \in
I^{(M)}_3$, then $\nu_4$ cannot be emitted in either $M^+_{\ell 2}$ or
$M^+_{\ell' 2}$ decay.

If for a given $m_{\nu_4}$, one knows, e.g., from peak-search experiments, that
$|U_{\ell' 4}|^2$ is sufficiently small that the denominator of (\ref{rr}) can
be approximated well by 1, then an upper bound on the deviation of 
$\bar R^{(M)}_{\ell/\ell'}$ from 1 yields an upper bound on $|U_{\ell 4}|^2$.
Thus, one has the bound
\beq
|U_{\ell 4}|^2 < \frac{\bar R^{(M)}_{\ell/\ell'}-1}
{\bar\rho(\delta^{(M)}_\ell,\delta^{(M)}_{\nu_4})-1} \quad {\rm for}
\ m_{\nu_4} \in I^{(M)}_2 \ .
\label{uellsqbound_i2}
\eeq
This gives very stringent upper limits on $|U_{\ell 4}|^2$ because
$\bar\rho(\delta^{(M}_e,\delta^{(M)}_{\nu_4}) >> 1$ over much of the interval 
$I^{(M)}_2$ (see Figs. 3-5 in \cite{shrock81a}).  If $m_{\nu_4} \in 
I^{(M)}_3$, then (\ref{rr}) reduces to $\bar R^{(M)}_{\ell/\ell'}=
(1-|U_{\ell 4}|^2)/(1-|U_{\ell' 4}|^2)$, so if 
$|U_{\ell' 4}|^2 << 1$ in this interval, then the upper limit is
\beq
|U_{\ell 4}|^2 < 1-\bar R^{(M)}_{\ell/\ell'}  \quad {\rm for}
\ m_{\nu_4} \in I^{(M)}_3 \ .
\label{uellsqbound_i3}
\eeq

We now apply this analysis to $R^{(\pi)}_{e/\mu}$, using (\ref{uellsqbound_i2})
and (\ref{uellsqbound_i3}) with $M^+=\pi^+$, $\ell=e$, and $\ell'=\mu$. From
previous $\pi^+_{\mu 2}$ peak search experiments
\cite{abela81}-\cite{triumf_pimu2} and the calculation of
$\bar\rho(\delta^{(\pi)}_\mu,\delta^{(\pi)}_{\nu_4})$, it follows that $|U_{\mu
  4}|^2$ is sufficiently small for $m_{\nu_4} \in I^{(\pi)}_2$ that we can
approximate the denominator of Eq. (\ref{rr}) by 1.  From 
$\bar R^{(\pi)}_{e/\mu}$ in Eq. (\ref{rcal_pi_emu}), using the procedure from
\cite{feldman-cousins}, we obtain the limit $\bar R^{(\pi)}_{e/\mu} <
1.0014$. Then, for $\nu_4 \in I^{(\pi)}_2$, we find
\beq
|U_{\ell 4}|^2 < \frac{\bar R^{(\pi)}_{e/\mu'}-1}
{\bar\rho(\delta^{(\pi)}_e,\delta^{(\pi)}_{\nu_4})-1}
<  \frac{0.0014}{\bar\rho(\delta^{(\pi)}_e,\delta^{(\pi)}_{\nu_4})-1} .
\label{uellsqbound_i2pi}
\eeq
This bound is labelled as PIENU in Fig. \ref{Ue4_figure}.  If $m_{\nu_4} \in
I^{(\pi)}_3$, i.e., $m_{\nu_4} > 139$ MeV, then, using (\ref{uellsqbound_i3}),
we obtain the upper bound on $|U_{e4}|^2$ given by the flat line labelled
PIENU-H in Fig. \ref{Ue4_figure}.

We next obtain a bound on $|U_{e 4}|^2$ by applying the same type of analysis
to $R^{(K)}_{e/\mu}$. From $K_{\mu 2}$ peak search
experiments \cite{asano81,bnl949,na62_2018} and the calculation of
$\bar\rho(\delta^{(K)}_\mu,\delta^{(K)}_{\nu_4})$, $|U_{\mu 4}|^2$
is sufficiently small that we can approximate the denominator of Eq. (\ref{rr})
well by 1.  Using Eq. (\ref{rcal_k_emu}) for $\nu_4 \in I^{(K)}_2$, we find
\beq
|U_{\ell 4}|^2 < \frac{\bar R^{(K)}_{e/\mu'}-1}
{\bar\rho(\delta^{(K)}_e,\delta^{(K)}_{\nu_4})-1}
<  \frac{0.010}{\bar\rho(\delta^{(K)}_e,\delta^{(K)}_{\nu_4})-1} .
\label{uellsqbound_i2k}
\eeq
This upper limit on $|U_{e4}|^2$ is labelled KENU in Fig. \ref{Ue4_figure}. 
For $m_{\nu_4} \in I^{(K)}_3$, i.e., $m_{\nu_4} > 493$ MeV, using 
({\ref{uellsqbound_i3}), we obtain the flat upper bound labelled 
KENU-H in Fig. \ref{Ue4_figure}.

One can also apply these methods to two-body leptonic decays of heavy-quark
hadrons. We first consider $D_s^+ \to \ell^+\nu_\ell$ decays \cite{cc}, using
(\ref{uellsqbound_i2}) and (\ref{uellsqbound_i3}) with $M^+=D_s^+$, $\ell=e$,
and $\ell'=\tau$.  Experimental data from CLEO, BABAR, Belle, and BES have
determined $BR(D_s^+ \to \mu^+ \nu_\mu) = (5.49 \pm 0.17) \times 10^{-3}$ and
$BR(D_s^+ \to \tau^+ \nu_\tau)=(5.48 \pm 0.23) \times 10^{-2}$
\cite{alexander09}-\cite{ablikim19}. Furthermore, searches by CLEO
\cite{alexander09}, BABAR \cite{babar10}, and Belle \cite{zupanc13} have
yielded the limit $BR(D_s^+ \to e^+ \nu_e) < 0.83 \times 10^{-4}$. Hence,
$R^{(D_s)}_{e/\tau} < 1.6 \times 10^{-3}$.  For $R^{(D_s)}_{e/\tau}$, using the
results of \cite{ms93}, we calculate $1+\delta_{RC}=0.948$. Substituting this
in Eq. (\ref{rmsm}) with $M=D_s$, $\ell=e$, $\ell'=\tau$, we find
\beq
R^{(D_s)}_{e/\tau,SM} = 2.29 \times 10^{-6} \ .
\label{rds_emu_sm}
\eeq
Therefore, $\bar R^{(D_s)}_{e/\tau} < 7.0 \times 10^2$. For
$R^{(D_s)}_{e/\tau}$, the interval $I^{(D_s)}_2$ is $191 \ {\rm MeV} <
m_{\nu_4} < 1.457$ GeV.  Actually, we restrict $m_{\nu_4}$ to a lower-mass
subset of this interval, because for sufficiently great $m_{\nu_4}$, even
though the $D_s^+ \to e^+ \nu_4$ decay is kinematically allowed to occur, the
momentum $p_e$ (in the $D_s$ rest frame) would be below the minimal value
set by experimental cuts in the BES III event reconstruction. With
$p_{e,cut} \simeq 0.8$ GeV \cite{bes3_pc}, this means that $m_{\nu_4}$ must be
less than 0.85 GeV for the event to be accepted. Thus, we consider $0.191 \
{\rm GeV} \ < m_{\nu_4} < 0.85$ GeV. Substituting the experimental limit on
$\bar R^{(D_s)}_{e/\tau}$ in the special case of (\ref{rr}) with $M=D_s$,
$\ell=e$, $\ell'=\tau$ and using the fact that $|U_{\tau 4}|^2 << 1$ for this
$m_{\nu_4}$ mass range \cite{pdg2018}, we obtain a resultant limit from
(\ref{uellsqbound_i2}). For $m_{\nu_4}=0.191$ GeV,
$\bar\rho(\delta_e^{(D_s)},\delta^{(D_s)}_{\nu_4}) = 1.37 \times 10^5$,
increasing to $\bar\rho(\delta_e^{(D_s)},\delta^{(D_s)}_{\nu_4}) = 1.83 \times
10^6$ for $m_{\nu_4}=0.85$ GeV.  We thus obtain the upper bound on $|U_{e4}|^2$
labelled $D_{se2}$ in Fig. \ref{Ue4_figure}.

A dedicated peak-search experiment to search for the heavy-neutrino decay
$D_s^+ \to e^+ \nu_4$
would be worthwhile and could improve the upper bound on
$|U_{e4}|^2$. Similarly, a search for leptonic $D$ decays like $D^+ \to e^+
\nu_4$ would be valuable and will be discussed elsewhere \cite{DB_RS}. The very
large values of $\bar\rho(\delta_e^{(D_s)},\delta^{(D_s)}_{\nu_4})$ and
$\bar\rho(\delta_e^{(D)},\delta^{(D)}_{\nu_4})$ over a large portion of the
kinematically allowed ranges of $m_{\nu_4}$ in $D_s^+ \to e^+ \nu_4$ and $D \to
e^+ \nu_4$ mean that there would be quite strong kinematic enhancement of the
heavy neutrino decay relative to the corresponding $(D_s^+)_{e2}$ and
$D^+_{e2}$ decays. In particular, these searches could be performed by the BES
III experiment, which recently reported results from a data sample of 3.19
fb$^{-1}$ and expects to collect considerably higher statistics.

Finally, we consider $B^+ \to \ell^+ \nu_\ell$ decays. There is an upper limit
$BR(B^+ \to e^+\nu_e) < 0.98 \times 10^{-6}$ from Belle \cite{satoyama07} and
BABAR \cite{aubert09}. For the other two leptonic decay modes, $BR(B^+
\to \mu^+ \nu_\mu) = (6.46 \pm 2.22_{stat} \pm 1.60_{syst} ) \times 10^{-7}$
from Belle \cite{belle_bmunu2018}, with a recent update $BR(B^+ \to \mu^+
\nu_\mu) = (5.3 \pm 2.0_{stat} \pm 0.9_{syst}) \times 10^{-7}$
\cite{belle_moriond2019,moriond2019}, and $BR(B^+ \to \tau^+ \nu_\tau) = (1.09
\pm 0.24) \times 10^{-4}$ from BABAR \cite{lees13} and Belle
\cite{hara13,kronenbitter15}.  The measured values of $BR(B^+ \to
\mu^+\nu_\mu)$ are in agreement with the SM prediction $BR(B^+ \to \mu^+
\nu_\mu)_{SM} = (3.80 \pm 0.31) \times 10^{-7}$ \cite{belle_bmunu2018}. The
measured value of $BR(B^+ \to \tau^+ \nu_\tau)$ is also in agreement with the
SM prediction $BR(B^+ \to \tau^+ \nu_\tau)_{SM} = (0.75^{+0.10}_{-0.05}) \times
10^{-4}$ \cite{kronenbitter15,ckmfitter}.

 We focus on data from a $B^+_{\ell 2}$ peak search
experiment by Belle \cite{park2016}. In general \cite{shrock80}, 
\beq
\frac{BR(M^+ \to \ell^+ \nu_4)}
     {BR(M^+ \to \ell^+ \nu_\ell)_{SM}} =
\frac{|U_{\ell 4}|^2 \bar\rho(\delta_\ell^{(M)},\delta_{\nu_4}^{(M)})}
     {1-|U_{\ell 4}|^2} \ .
\label{br_ratio}
\eeq
For $m_{\nu_4}$ in the range from 0.1 GeV to 1.4 GeV, the Belle experiment
obtained an upper limit on $BR(B^+ \to e^+ \nu_4)$ of $2.5 \times 10^{-6}$,
while in the interval of $m_{\nu_4}$ from 1.4 GeV to 1.8 GeV, this upper limit
increased to $7 \times 10^{-6}$. In the range of $m_{\nu_4}$ from 0.1 to 1.3
GeV, the Belle experiment obtained (non-monotonic) upper limits on $BR(B^+ \to
\mu^+ \nu_4)$ of approximately $2-4 \times 10^{-6}$, and in the interval of
$m_{\nu_4}$ from 1.3 GeV to 1.8 GeV, it obtained upper limits varying from $2
\times 10^{-6}$ to $1.1 \times 10^{-5}$. Substituting the $BR(B^+ \to
e^+\nu_4)$ limits in Eq. (\ref{br_ratio}) with $M=B$ and $\ell=e$, we obtain
the upper limits on $|U_{e4}|^2$ shown as the curve $B_{e2}$ in
Fig. \ref{Ue4_figure} \cite{banomaly}. From the $BR(B^+ \to e^+\nu_4)$ limits
we infer upper limits on $|U_{\mu 4}|^2$ that decrease from 0.83 to $3.4 \times
10^{-2}$ as $m_{\nu_4}$ increases from 0.1 GeV to 1.2 GeV.
Further peak searches for $B^+ \to \ell^+ \nu_4$ with $\ell=e, \ \mu$ 
at Belle II would be worthwhile as a higher-statistics extension of 
\cite{park2016}. 

We briefly remark on other constraints on a Dirac $\nu_4$ in the mass range
considered here. From the results of \cite{p77,leeshrock77}, it follows that
there is a negligibly small contribution to decays such as $\mu \to e
\gamma$ and $\mu \to ee\bar e$. Similarly, there is no conflict with bounds on
neutrino magnetic moments \cite{fs80,pdg2018}, and contributions to
invisible Higgs decays \cite{invis} are  well below the current upper limit of
$BR(H \to {\rm invis.}) < 19 \%$ \cite{lhc_invis}.

In this work, improved  upper limits on $|U_{e4}|^2$  
have been presented covering most of the range from
$m_{\nu_4} =0.5$ MeV to $m_{\nu_4} \simeq 1$ GeV, 
representing  the best available laboratory bounds for a Dirac
neutrino $\nu_4$ that do not make model-dependent assumptions concerning
visible neutrino decay modes. Over parts of this range, the bounds obtained
are competitive with those that assume specific visible $\nu_4$ decays.
For example, for $m_{\nu_4}=30$ MeV, our upper bound is $|U_{e4}|^2 < 0.8
\times 10^{-6}$, while the best bound for this value of $m_{\nu_4}$ from
experiments searching for neutrino decays is $|U_{e4}|^2 < 1 \times 10^{-6}$
\cite{ps191}.  New peak search experiments to search for $D_s^+ \to e^+ \nu_4$
and $D^+ \to e^+ \nu_4$ as well as a continued search for $B^+ \to e^+ \nu_4$
would be valuable; these could improve the bounds further.  Other constraints
on sterile neutrinos such as from $\pi^+\to\pi^0 e^+ \nu_e$
decay,and a detailed report of the results presented here will be 
published elsewhere \cite{DB_RS}.

We thank J. Benitez, J. Hardy, V. Luth, W. Marciano, M. Ramsey-Musolf, C. Yuan,
and G. Zhao for useful discussions.  This work was supported in part by the
Natural Sciences and Engineering Research Council and the National Research
Council of Canada (D.B.) and by the U.S. National Science Foundation Grant
NSF-PHY-16-1620628 (R.S.).



\begin{thebibliography}{99}
\bibitem{st}
We will use the term ``sterile neutrino'' both in its precise sense as 
an electroweak-singlet interaction eigenstate and in a commonly used
approximate sense as the corresponding mainly sterile mass eigenstate(s) 
in this neutrino interaction eigenstate.
\bibitem{vudapp}
%
  Since in the presence of massive neutrinos, $G_{F,app}$ would differ 
  from $G_F$, the value of $|V_{ud}|$, denoted $|V_{ud,app}|$, derived from 
  these fits to ${\cal F}t$ values would also differ from the true value
  \cite{shrock81a,shrock_vpi}. However, all that we use here is the relative
  agreement between the ${\cal F}t$ values, as measured by the quoted precision
  in the determination of $|V_{ud,app}|$, so we keep the subscript $app$
  implicit.
\bibitem{soft}
%
Sufficiently soft photons are understood to be included in the experimentally
measured $M^+ \to \ell^+ \nu_\ell$ decays, which are thus often labelled as 
$M^+ \to \ell^+ \nu_\ell (\gamma)$.
\bibitem{cc}
Charge-conjugate decays are understood to be included here and below.
\end{thebibliography}
\end{document}